\documentclass[twocolumn,prd,nofootinbib,aps,prl,floats,floatfix,superscriptaddress,amsmath,amssymb,longbibliography,secnumarab
ic]{revtex4-2} %

\usepackage[final]{graphicx}
\usepackage{hyperref}
\usepackage{amsmath}
\usepackage{bbm}
\usepackage{amsfonts}
\usepackage{amssymb}
\usepackage{latexsym}
\usepackage{graphicx}
\usepackage[english]{babel}
\usepackage{multirow}
\usepackage{float}
\usepackage{url}
\usepackage{slashed}
\usepackage{xcolor} 
\usepackage[utf8]{inputenc}
\usepackage{verbatim}
\usepackage{stackengine}
\usepackage{cancel}
\usepackage{accents}
\usepackage{array}
\usepackage{physics}
\usepackage{bm}

\newcommand{\be}{\begin{equation}}
\newcommand{\ee}{\end{equation}}
\newcommand{\ba}{\begin{array}}
\newcommand{\ea}{\end{array}}
\newcommand{\bea}{\begin{eqnarray}}
\newcommand{\eea}{\end{eqnarray}}

\usepackage{slashed}

\begin{document}
\rightline{FERMILAB-PUB-25-0730-T}
\title{
New Tests of Low-Scale Quantum Gravity with Cosmic-Ray Collisions}
\author{Manuel Ettengruber}
\email{manuel@mpp.mpg.de}
\affiliation{Universit{\'e} Paris--Saclay, CNRS, CEA, Institut de Physique Theorique, 91191, Gif-sur-Yvette, France
}
\author{Gonzalo Herrera}
\email{gonzaloh@mit.edu}
\thanks{ORCID: \href{https://orcid.org/0000-0001-9250-8597}{0000-0001-9250-8597}}
\affiliation{Kavli Institute for Astrophysics and Space Research, Massachusetts Institute of Technology, Cambridge, MA 02139, USA}

\affiliation{Harvard University, Department of Physics and Laboratory for Particle Physics and Cosmology, Cambridge, MA 02138, USA}
\affiliation{Center for Neutrino Physics, Department of Physics, Virginia Tech, Blacksburg, VA 24061, USA}

\begin{abstract}

Cosmic ray collisions at high center of mass energy could enable graviton and black hole production as expected in theories of low-scale quantum gravity, such as extra-dimensions, many species, or some versions of string theory. Here we propose three novel phenomenological tests of these theories. We first consider the collision of cosmic rays with ambient protons, electrons and photons in Active Galactic Nuclei (AGN), finding that high-energy neutrino data from the blazar TXS 0506+056 places a constraint on the fundamental scale of gravity of $M_{f} \gtrsim $ 0.3 TeV, and future high-energy neutrino data could raise this bound to $M_{f} \gtrsim 200$ TeV. We then point out that collisions of pairs of cosmic rays could occur at a sizable rate in AGN where the accelerated cosmic rays are not collimated, or on supermassive black hole binaries. This consideration could potentially let us test unprecedented large fundamental scales of $M_{f} \gtrsim$ 2 PeV. We further compute the corresponding thermal neutrino emission arising from the Hawking evaporation of black holes produced in cosmic ray collisions, finding a spectrum that clearly differs from that expected in meson decays. Finally, we speculate with an scenario which would produce high-energy neutrino and gamma-ray emission from regions in the sky where no multi-wavelength counterparts would be expected, via graviton propagation from a different brane, which then decays in our Universe.

\end{abstract}
\maketitle

\emph{\textbf{Introduction.}}\label{sec:introduction}

The question on what is the actual fundamental scale of gravity, $M_f$, has triggered renewed interest over the years. It has been long believed that $M_f$ coincides with the Planck scale $M_P$, but in \cite{Dvali:2007hz} it has been shown that the number of particle species, $N$, present within a theory, separates the scale where gravity actually becomes strong from the Planck scale according to the formula
\begin{equation}
    M_f = \frac{M_P}{\sqrt{N}}\;.
    \label{masterformula}
\end{equation}
The first theory that made use of this was the model of large extra dimensions fathered by Arkani-Hamed, Dimopoulos and Dvali (ADD) \cite{Arkani-Hamed:1998jmv}, where $M_f$ was brought down to a TeV-scale, providing a solution to the hierarchy problem. 

What this equation tells us is that the scale where quantum gravity becomes relevant is not around $M_P$ but around $M_f$. Processes that have a momentum transfer $q$ where $q \gg M_f$ will produce a black hole (BH) due to the concept of classicalization in this regime \cite{Dvali:2010jz, Dvali:2014ila}. The core of classicalization can be understood with the following \textit{Gedankenexperiment}: Imagine that two particles collide with $q \gg M_f$. Then the length scale of the process will be shorter than the Schwarzschild radius $ r_s \sim M_f^{-1}$ and the stored energy on this length scale will be greater than $M_f$. From General Relativity, we know that in such a situation a BH is formed, and in theories with $M_f \ll M_P$ one expects that this is possible on a much larger length scale (lower energy scale) than originally anticipated. 

When the momentum transfer is not much smaller than the cutoff $q \leq M_f$, the phenomenology stemming from the collective effect of the existence of $N$ species does scale as $q/M_f$ and cannot be neglected \cite{Arkani-Hamed:1998sfv, Dvali:2009ne}. Depending on the model at hand the resulting phenomenological consequences can be very different,, but some features are shared by all kinds of such models. 

\begin{figure*}[t!]
\centering
\includegraphics[width=0.8\textwidth]{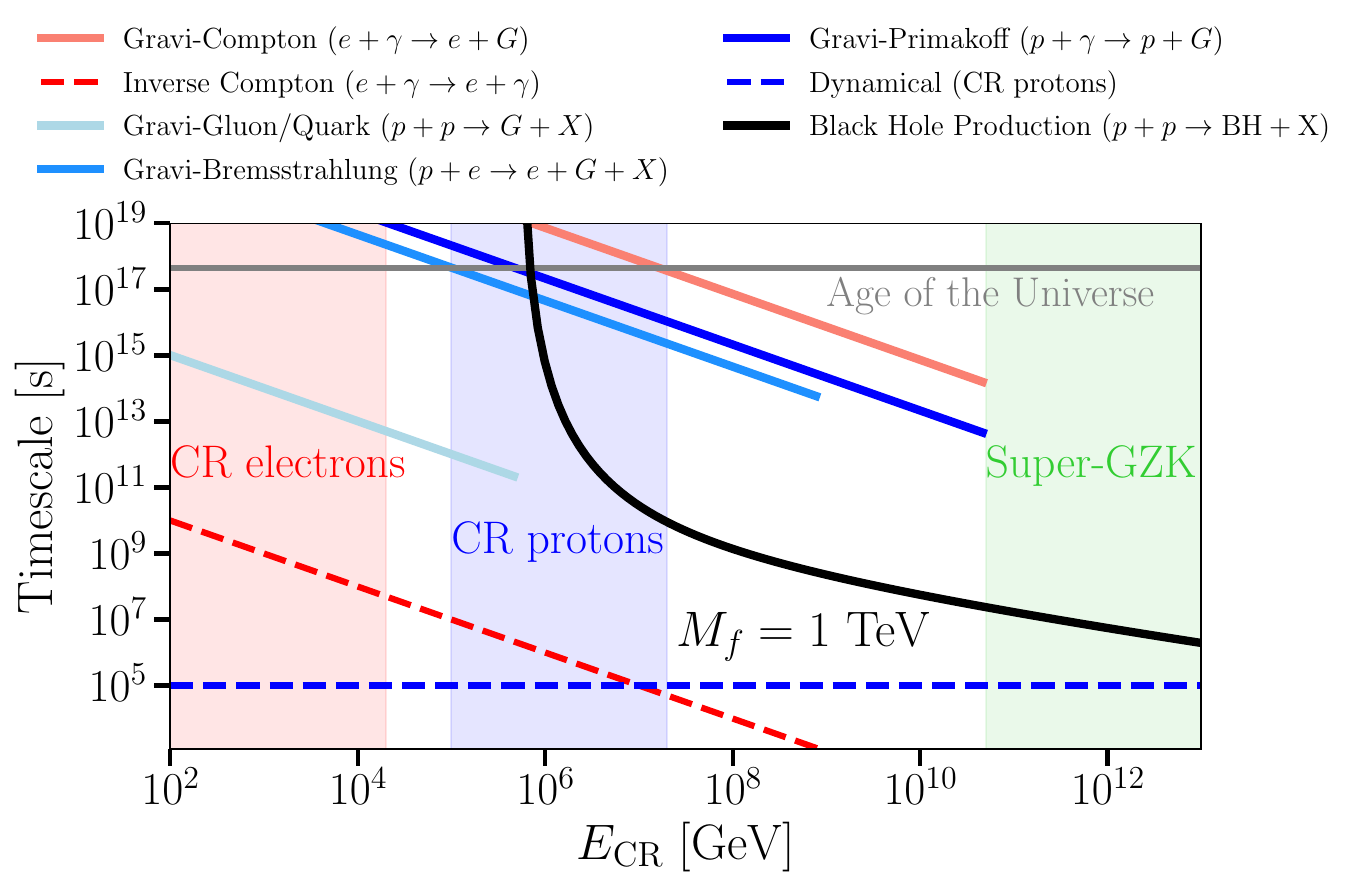}
\caption{Timescales for Graviton and Black Hole production from cosmic ray processes in TXS 0506+056, for a new fundamental scale of $M_{f}=1$ TeV and $n=2$. The cut-offs in the timescales corresponds to the limit where the EFT description of graviton processes becomes invalid and black hole production begins to occur, at $\sqrt{s}=M_{f}$. This cut-off is different depending on the process considered. We confront these timescales with the inferred dynamical timescale of cosmic ray protons from TXS 0506+056, and the timescale from inverse compton scattering of cosmic ray electrons. This serves as a benchmark for the leading Standard Model processes cooling cosmic rays in TXS 0506+056. We further show vertical shaded bands indicating the energies at which we have experimental evidence from cosmic ray electron and proton acceleration. We also highlight the region where cosmic rays would be accelerated somewhat beyond the GZK limit, to indicate the maximum cosmic ray energies plausibly achieved in a source of this characteristics. This will be used to set projected constraints (see main text for details).}
\label{fig:timescales_TXS}
\end{figure*}

Most of the models that exploit Eq. \eqref{masterformula} were built to solve the hierarchy problem of particle physics \cite{Arkani-Hamed:1998jmv, Dvali:2008fd, Arkani-Hamed:2016rle}, but it can be also used to explain the Cosmological Constant problem \cite{Montero:2022prj}, the smallness of neutrino masses \cite{Arkani-Hamed:1998wuz, Dvali:1999cn, Dvali:2008fd, Anchordoqui:2022svl, Ettengruber:2025usk, Ettengruber:2022pxf}, and provide suitable candidates for the dark matter of the Universe \cite{Arkani-Hamed:1998sfv, Arkani-Hamed:1999rvc, Dvali:2008fd, Arkani-Hamed:2016rle, Gonzalo:2022jac, Friedlander:2022ttk, Anchordoqui:2024dxu, Anchordoqui:2024jkn, Anchordoqui:2025opy, Ettengruber:2025kzw}. 

These theories have received significant attention in the community, but with LHC reaching its maximum center of mass energy its phenomenological consequences and possible testability with particle collisions not being sourced by colliders have not been explored in much depth yet.

Naturally, if aiming to reach higher fundamental scales $M_f$, we need to look at the most energetic environments in the Universe. One class of environments for which processes with energetics beyond the TeV must occur are Active Galactic Nuclei (AGN). The IceCube collaboration presents evidence of high-energy neutrino emission from two of these sources, TXS 0506+056 \cite{IceCube:2018dnn}, and  NGC 1068 \cite{IceCube:2022der} \footnote{Recently, the KM3Net collaboration observed a neutrino at larger energy, but this has yet not been conclusively linked to any electromagnetic counterpart \cite{KM3NeT:2025npi, Fang:2025nzg}.}. Plausibly, these high-energy neutrinos originate from cosmic ray protons that have been accelerated to very high energies within the AGN, then scattering efficiently with ambient protons and/or photons in the AGN. Within the scales of the AGN, cosmic rays can be accelerated even beyond the GZK limit. For instance, the center of mass energy of a GZK proton scattering on a target proton at rest is 
\begin{equation}
\sqrt{s_{\rm GZK, pp}} \sim \sqrt{2m_pE_{p}} \sim 300 \, \mathrm{TeV}
\end{equation}
which is well beyond collider probes. Conceivably, the maximum achievable center of mass energy in the AGN scales due to cosmic ray collisions with ambient protons could even be larger, since the GZK limit only applies to ultra-high energy cosmic ray protons propagating through the CMB over cosmological distances.
The high-energy neutrino observations from NGC 1068 and TXS 0506+056 have been exploited to set limits on some Beyond the Standard Model scenarios, see \textit{e.g} \cite{Wang:2021jic,Cline:2022qld, Ferrer:2022kei, Cline:2023tkp, Herrera:2023nww,Herrera:2025gpm,Carloni:2022cqz, Hyde:2023eph, Ciscar-Monsalvatje:2024tvm, Zapata:2025huq, DeMarchi:2024riu, Wang:2025ztb, Gustafson:2025dff, Bhowmick:2022zkj}, but not on low-scale quantum gravity theories. These theories are particularly suited to be probed with cosmic rays produced in AGN, due to their manifestation at high-center of mass energies.

In this paper we will focus on TXS 0506+056, since the high-energy neutrino emission was more energetic than NGC 1068. The interactions of cosmic ray protons with ambient protons in these sources could have center of mass energies that go beyond the fundamental scale of gravity $\sqrt{s} \geq M_f$, therefore leading to gravitational induced processes that could cool cosmic rays in AGN more efficiently than ordinary SM processes. This consideration, properly quantified and justified given our current understanding of AGN emission and energetics, can be used to place new lower limits on the fundamental scale of gravity $M_f$.

In addition, we also propose a novel mechanism to test low-scale gravity theories, relying on cosmic ray collisions with other cosmic rays. This process could be particularly relevant in environments where the cosmic ray acceleration is not collimated, for instance stochastic acceleration or turbulence in AGN cores. The maximum center of mass energy in this case can be very large. For instance, two GZK cosmic rays colliding yield

\begin{equation}
\sqrt{s_{\rm GZK, pp}} \sim \sqrt{2}E_p \sim 7\times 10^{4} \, \mathrm{PeV} \sim 6 \times 10^{-9} M_P.
\end{equation}
Potentially, in some AGNs, this number could be even larger, since the GZK limit only applies on extragalactic propagation scales. However, the fact that these large c.o.m energies can be achieved does not mean that these scales are testable. It is also important to assess whether these processes could be efficient enough to yield observable signatures. We will address in some detail such a possibility in this note, focusing on AGNs and our current understanding on these environments.

\begin{figure*}[t!]
\centering
\includegraphics[width=0.7\textwidth]{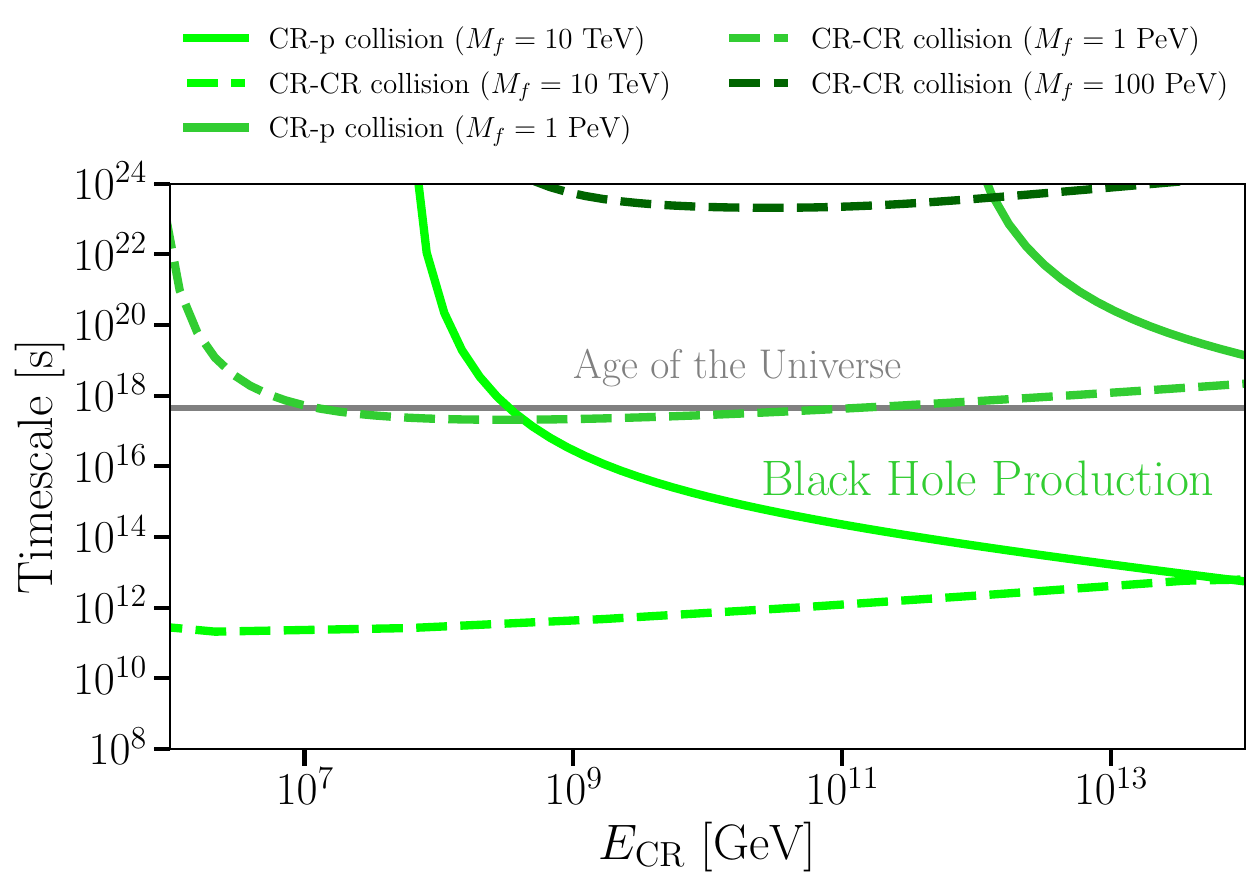}
\caption{Timescales for black hole production via cosmic ray collisions with ambient protons (solid) and off other cosmic rays (dashed), for $n=2$. The different colors correspond to different scales of low-scale gravity $M_{f}$. For low-energy ambient protons, a number density of $n_p=10^{7}$cm$^{-3}$ is assumed. For the cosmic ray-cosmic ray collisions, a magnetic field of $B=100$ kG and acceleration efficiency of $\eta=1$ is taken. For comparison, we show the age of the Universe, which indicates a conservative upper limit on the maximum timescale that could conceivably lead to observable signatures.}
\label{fig:timescales_UHECR}
\end{figure*}
\emph{\textbf{Graviton and black hole production from cosmic ray interactions.}}\label{sec:theory}
In low-scale gravity theories, both gravitons and black holes can be produced from processes undergoing SM particles. In this work we will consider a variety of processes that can occur in AGN, and are able to yield gravitons $G$ or black holes $\mathrm{BH}
$ in the final state.

First, the dominant process for cosmis ray electrons is cross section for Gravi-Compton scattering with ambient low-energy photons $\gamma +e \rightarrow e+G$ reads \cite{Arkani-Hamed:1998sfv, Giudice:1998ck}
\begin{equation}
    \sigma \sim \frac{e^2}{16\pi} \frac{E^n}{M_{f}^{2+n}}.
\end{equation}

The analogous process for cosmic ray protons (or nuclei with atomic mass Z) is Gravi-Primakoff scattering, with cross section $(N)p+\gamma \rightarrow p+G$
\begin{equation}
    \sigma \sim \frac{Z^2}{16\pi}\frac{E^n}{M_f^{n+2}}
\end{equation}
Cosmic ray protons (or nuclei) can also scatter with ambient electrons (and viceversa) via Gravi-Bremsstrahlung $p(N)+e \rightarrow e+G+X$
\begin{equation}
\sigma \sim \frac{Z^2 e^2}{16 \pi} \frac{E^n}{M_{f}^{n+2}}
\end{equation}

Another relevant process from pp collisions  is via Gravi-Gluon and Gravi/Quark production. the cross section for $qq \rightarrow g + G$ can be determined by the ratio
\begin{equation}
    \frac{\sigma_{ee\rightarrow \gamma G}}{\sigma_{qq\rightarrow g G}} \simeq \frac{\frac{\alpha}{16}}{\frac{\alpha_s}{36}}.
\end{equation}

We will consider all the aforementioned processes yielding the production of gravitons in our analysis.



Turning to black hole production in low scale gravity theories we notice that a BH can be formed if the impact parameter of two particles with $\sqrt{s}> M_f$ is smaller than the Schwarzschild radius, which is given in such theories by \cite{Myers:1986un, Argyres:1998qn, Dvali:2008fd}
\begin{equation}
    r_s = \frac{a_n}{M_f} \left(\frac{M_{\rm BH}}{M_f} \right)^\frac{1}{1+n} \; ,
    \label{eq:Schwarzschildsemiclassical}
\end{equation}
where $n$ and $a_n$ are model-dependent parameters. For extra-dimensional theories the parameter $n$ can be identified with the number of extra-dimensions and 
\begin{equation}
    a_n = \left(
                8\mspace{1mu}\pi^{-(n-1)/2}
                \frac{
                \Gamma
                \big[
                    (n+3)/2\mspace{1mu}
                \big]}{n+2}
            \right)^{\!1/1+n}
            \, .
\end{equation}
In many species theories, $a_n=1$ and $n>2$.

Therefore, the resulting cross section for the formation of a BH is the geometric cross section which is given by
\begin{equation}
    \sigma_0 = \pi r_{s}^2 \;.
    \label{productioncrosssection}
\end{equation}
Such a BH has a mass that is larger than $M_f$, but will not be an Einsteinian BH's. This is intuitively easy to understand by analyzing the two extreme mass regimes of BH. As $M_f$ signals the scale where Quantum Gravity is dominant, BH's of such a mass will be quantum states. On the other hand, BH's of astrophysical masses are fully Einsteinian. Therefore, a transition period between these two regimes is unavoidable, and in \cite{Dvali:2008fd, Dvali:2008rm} this regime has been given for the species case as $M_f \ll M_{\rm BH} \ll \sqrt{N}M_P$. For an extra-dimensional scenario, the beginning of the transition regime has a geometric interpretation, namely when the size of the BH gets smaller than the compactification radius, which means the BH is then correctly described by an higher-dimensional one instead of a 4-dimensional BH. In this transition regime, gravity is still weak, but the properties of the BH deviate from an Einsteinian one, like the change of the Schwarzschild radius as discussed above. 

In reality, at sufficiently high c.o.m. energy, the $pp$ collisions occur in the deep inelastic regime. In this case, to properly describe the cross section, we compute \cite{Barrau:2005zb}
\begin{align}
\frac{d \sigma_{p p \rightarrow \mathrm{\rm BH}+X}}{d M_{\rm BH}} 
&= \left. \sigma_0 \right|_{\sqrt{s}=M_{\rm BH}} \frac{2 M_{\rm BH}}{s} \notag \\
&\quad \times \sum_{i, j} \int_{M_{\rm BH}^2 / s}^1 \frac{d x_i}{x_i} 
f_i\left(x_i\right) f_j\left(\frac{M_{\rm BH}^2}{s x_i}\right) \; ,
\end{align}
where $f_a(x_a)$ are the parton distribution functions (PDFs) encoding the probability of finding a parton of type $a$ carrying a fraction $x_a$ of the proton momentum at a given factorization scale. We approximate the gluon PDF as extracted from \texttt{LHAPDF} as \cite{Butterworth:2015oua}
\begin{equation}
f_g(x) \approx x^{-1.3}(1 - x)^5.
\end{equation}
For light quarks and antiquarks, we adopt the forms
\begin{equation}
f_q(x) \simeq x^{-0.5}(1 - x)^3, \qquad f_{\bar{q}}(x) \simeq x^{-0.5}(1 - x)^5,
\end{equation}
which qualitatively reproduce the typical shape of valence and sea quark distributions. We include all leading partonic channels in the cross section, summing over $gg, qg,qq, q\bar{q}$ initial states, with appropriate multiplicities for the number of active quark flavors. The total cross section for black hole production in $pp$ collisions then reads
\begin{equation}
\sigma_{p p \rightarrow \mathrm{BH}+\mathrm{X}}(s)=\int_{M_{f}}^{\sqrt{s}} d M_{\mathrm{BH}} \frac{d \sigma_{p p \rightarrow \mathrm{BH}+X}}{d M_{BH}}\; .
\end{equation}

As the Schwarzschild radius of semiclassical BH's changes according to \eqref{eq:Schwarzschildsemiclassical} the Hawking temperature for a theory with $n$ extra-dimensions is 
\begin{equation}
    T_H =\frac{n + 1}{4 \pi r_s}\; .
    \label{HawkingTempADD}
\end{equation}
By plugging \eqref{eq:Schwarzschildsemiclassical} into this equation we see that BH's with masses $M_f$ will also radiate with a temperature of $M_f$ (similar for the many species case as $T_H = r_s^{-1}$ ). Because the lifetime of such semiclassical BH's is short, one could think that this channel may still lead to a sizable gamma-ray and neutrino emission in AGN, similarly to SM processes, rendering the model untestable at first glance. However, this intuition is wrong for a few reasons.
First, such BH's do not evaporate simply back into the particles that produced them, but gravitational processes can go into all thermally allowed species. In extra-dimensional models a significant amount of produced particles by the Hawking radiation will be gravitons \cite{Cardoso:2005mh, Friedlander:2022ttk} if the number of extra dimensions is large enough. These Gravitons will have dominantly the mass of $T_H$ which is again maximally of order $M_f$ and can be potentially be longlived according to \cite{Arkani-Hamed:1998sfv}
\begin{equation}
    \tau \approx 10^{10} \mathrm{yr} \times\left(\frac{100 \mathrm{MeV}}{T_H}\right)^3 \approx 10^{-3} \mathrm{yr}  \;.
\end{equation}

This effect is even more dramatic in the many species scenario where the black hole can evaporate in all thermally allowed dark species. The fraction of energy release between our species and all available species can be calculated as \cite{Dvali:2008fd}
\begin{equation}
    \frac{E_{\rm BH \rightarrow \textrm{our copy}}}{E_{\rm BH \rightarrow \textrm{other copies}}} = \left( \frac{M_f}{M_{\rm BH}} \right)^\frac{n}{n+1} \;.
\end{equation}

The second effect and even more importantly as it acts in both models equally is that the produced particles from Hawking evaporation have much lower temperature as the original particles that formed the black hole in the first place. In order to demonstrate this let us investigate a typical scenario with two extra dimensions, so $n = 2$ and $M_f = 10 \textrm{ TeV}$. In Table \ref{tableTH} we show the Hawking temperature for BH's stemming from different center of mass energies, and calculate the ratio of the energy of the product particles and the inital particles.

\begin{table}[h!]
\centering
 \begin{tabular}{||c c c||} 
 \hline
 $\sqrt{s}$ in TeV& $T_H$ in TeV & $T_H/\sqrt{s}$\\ [0.5ex] 
 \hline\hline
 50 & 1.4 & 0.03 \\ 
 100 & 1.1 & 0.01 \\
 300 & 0.8 & $2.7\times10^{-3}$ \\
[1ex] 
 \hline
 \end{tabular}
 
 \caption{For a scenario with $n=2$ and $M_f = 10 \textrm{ TeV}$ the c.o.m. energy of the production process is shown in the left column, the resulting BH Hawking temperature in the middle, and the ratio between the temperature and the initial energy in the right column.}
\label{tableTH} 
\end{table}

One can notice that the temperature of the products is lowered at least by two orders of magnitude, and this effect becomes stronger for more massive black holes. This means that as more energetic the initial protons are, the stronger the cooling is. 

The third reason why we expect BH cooling to be efficient in this regime is that the assumption of full evaporation may not hold. Just by the nature of Hawking radiation, which is a semi-classical effect, it is clear that the way BH's evaporate cannot be extrapolated arbitrarily at all scales. Until recently, the agreement within the field is that after the page time \cite{Page:1993wv}, BH evaporation should change drastically from the ordinary Hawking picture. Whatever happens at this point is unclear now, but if the dying black hole is realizing its energy back into the jet it would do so with maximally at a temperature of $M_f$ and therefore would not cool the jet further.

Nevertheless, other proposals have been made. Recent developments around the Memory burden effect on BH's suggest that they should get stabilized at latest if they lost half of their mass \cite{Dvali:2018xpy, Dvali:2019jjw, Dvali:2019ulr, Dvali:2020wqi, Dvali:2020wft,Ettengruber:2025kzw}. If such a stabilization takes place, BH formation acts as an energy sink in AGN.

In the remaining of this letter we will study the implications of graviton and black hole production in AGNs. High-energy neutrino observations, combined with gamma-rays, allow us to infer the cosmic ray luminosity and spectra from some extragalactic Active Galactic Nuclei, and the corresponding cooling timescales via Standard Model processes. Of particular interest is the source TXS 0506+056, the first ever detected extragalactic high-energy neutrino source, also bright in very high-energy gamma rays \cite{IceCube:2018dnn}. It has been discussed that cosmic rays in TXS 0506+056 do not cool efficiently, and that the fate of cosmic rays is governed by their dynamical timescale \cite{Keivani:2018rnh}. This is simply given by $\tau_{\rm dyn} \sim R/c$ with $R$ the size of the radius of the blob where cosmic ray acceleration occurs, and $c$ is the speed of light. For TXS-0506+056, its value is $\tau_{\rm dyn} \sim R_{\rm jet}/c \sim  10^{5}$s in the blob frame, which turns out smaller than the cooling timescales induced by photopion and proton-proton collisions at sufficiently high energies \cite{Keivani:2018rnh}.

The induced cosmic-ray cooling timescales from $pp \rightarrow X+\mathrm{BH}$ and $pp \rightarrow G+X$ processes is given by

\begin{equation}
\tau_{pp}=\frac{1}{n_p \sigma_{pp}c}.
\end{equation}

where $n_p$ is the number density of protons in TXS 0506+056. This value is uncertain, depending on the astrophysical model under consideration. Here we assume $n_p =10^{12} \mathrm{cm}^{-3}$ \cite{Yang:2024bsf, Fiorillo:2025cgm}.

In Figure \ref{fig:timescales_TXS}, we show the timescales induced by graviton and black hole production from cosmic ray protons (blue lines) and cosmic ray electrons (red lines), for different values of $n$ and a fixed scale of $M_{f}=1$ TeV. For comparison, we show the dynamical timescale of cosmic ray protons, the timescale from inverse compton scatterings of cosmic ray electrons, the age of the Universe, and colored bands for the observed cosmic ray cooling regions from TXS 0506+056 neutrino and gamma-ray data \cite{IceCube:2018cha, Keivani:2018rnh, Herrera:2023nww}. It is noticeable that the induced timescales from graviton and black hole production at $M_f =1$ TeV are larger than those from Standard Model processes at the relevant energies, however, lower scales could compete with the dynamical and inverse compton timescales.

We also notice that our plots extend only up to $E_p \simeq 10^{12}$ GeV, only mildly beyond the GZK limit. There is no evidence for cosmic ray acceleration going much farther beyond this energy. It is also important to discuss the dependence of the cooling timescales with $n$ ad $M_f$. While the value of $M_f$ only affects the overall normalization of the cross section for black hole production, the number of extra-dimensions $n$ affects the dependence of the cross section with the cosmic ray energy (or black hole mass), confer Eq. \ref{eq:Schwarzschildsemiclassical}. In particular, the dependence of the cross section with energy becomes milder for larger value of $n$, which translates into larger timescales beyond some cosmic ray energy threshold. However, for sufficiently low cosmic ray proton energies, larger values of $n$ lead to shorter timescales.

We will derive upper limits on low-gravity theories from the consideration that cosmic rays in TXS 0506+056 shall not cool much faster than is inferred from high-energy neutrino and gamma-ray data, that is than a dynamical timescale of $\sim 10^5$s. Concretely, we place a limit on the combination of $n$ and the new fundamental scale $M_f$ as

\begin{equation}\label{eq:timescales_limit}
\text{max}(\tau_{pp \rightarrow X+BH}, \tau_{pp \rightarrow G+X}) \gtrsim C \tau_{\rm dyn} 
\end{equation}
where we take $C \sim 0.1$ , since the cosmic ray luminosity inferred from high-energy neutrino data from TXS 0506+056 already violates the Eddington luminosity \cite{IceCube:2018cha,Murase:2018iyl}.

\begin{figure*}[tb]
\centering
\includegraphics[width=0.75\textwidth]{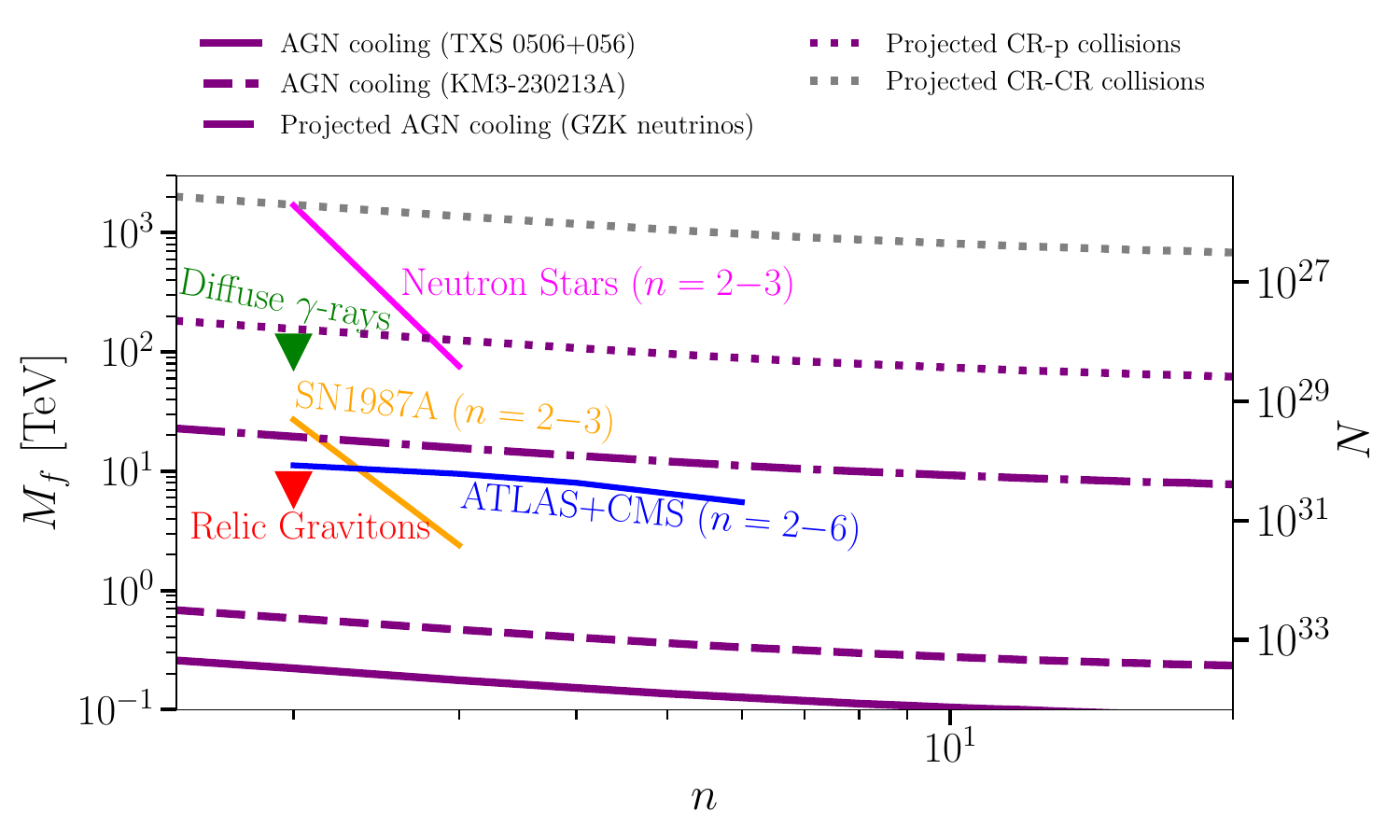}
\caption{Upper limits on the fundamental scale of gravity $M_f$, or number of species N, versus the parameter $n$, identified with the number of extra-dimensions in some theories. For comparison, we also show upper limits from graviton production in SN1987A \cite{Hanhart:2001fx} and cosmological bounds from graviton decay contributions to the diffuse gamma-ray flux. It has been discussed that these bounds are relaxed if gravitons can also decay on other branes \cite{Hall:1999mk}. We also show collider bounds from \cite{ATLAS:2021kxv,CMS:2017zts}.}
\label{fig:limits}
\end{figure*}

In Figure \ref{fig:limits} we show in a purple solid line the limit on $M_f$ (or equivalently, the number of many-species $N$) as a function of $n$ from TXS 0506+056, obtained from evaluating Eq. \ref{eq:timescales_limit} in the range $E_{\rm CR}=0.1-20$PeV. As a dashed purple line, we show plausible limits from the KM3-230213A event, were this event arising from an AGN source, and where the maximum cosmic ray energy assumed is $E_{\rm CR}=2000$ PeV. We further show projected limits arising from the potential future detection of GZK neutrinos as a dashed-dotted purple line, with $E_{\rm CR}=10^{11}$ GeV. Finally, we show as a dotted grey line the most conceivable stringent limit one could derive from cosmic ray-cosmic ray collisions. For comparison, we show a variety of past astrophysical and laboratory limits derived in this parameter space \cite{Hanhart:2001fx, Hall:1999mk, ATLAS:2021kxv,CMS:2017zts}.

\emph{\textbf{Cosmic ray-cosmic ray collisions}}

One may also consider the possibility that pairs of cosmic rays collide, which enhances the available center-of-mass energies compared to collisions of cosmic rays with protons at rest. This circumstance is very unlikely in blazars such as TXS 0506+056, where cosmic rays are accelerated in a collimated beam. However, in environments without jets—where cosmic rays are accelerated stochastically in turbulent magnetic fields and plasma waves—the particle distribution tends to isotropize \cite{1949PhRv...75.1169F, 1989ApJ...336..243S, 2012SSRv..173..535P}. Such isotropy significantly increases the likelihood of cosmic-ray–cosmic-ray collisions, in contrast to strongly beamed jet scenarios where particles move nearly parallel.  To assess the importance of this process, we first estimate the number density of cosmic rays compared to that one of thermal protons in the an AGN where cosmic rays are accelerated. As an example, one may consider an NGC 1068-like source, where the acceleration may be stochastic in origin and no strong jets have been observed \cite{Murase:2022dog}. To do so we first calculate the energy density that goes into cosmic rays, $U_{\rm CR}$, via
\begin{equation}
    U_{\rm CR} = \frac{\eta_{\rm eff}}{2} U_B = \frac{\eta_{\rm eff}}{2} B^2 \;,
\end{equation}
where $B$ is the magnetic field within the corona of the AGN and $\eta_{\rm eff}$ is the acceleration efficiency. Then one can translate the $U_{\rm CR}$ into a number density with the typical cosmic ray energy $E_{\rm CR}$ with
\begin{equation}
    n_{\rm CR} = \frac{U_{\rm CR}}{E_{\rm CR}} \;.
\end{equation}
Knowing this we can estimate the ratio with the thermal protons $n_{\rm CR}/n_{\rm p}$. To get a feeling of whether cosmic-ray collision events could be relevant, we take optimistic but plausible parameters for some cosmic ray accelerators, $B= 100$ kG \cite{Tchekhovskoy_2011,Piotrovich:2020ooz,Blanco:2023dfp, Blanco:2025zqo, Lisakov:2024jlp}, and $ \eta_{\rm eff} = 1$ \cite{Sironi:2015eoa, Sironi:2025kgn, Comisso:2018kuh}. Then we can rearrange the previous expression on the number density of cosmic rays as

\begin{equation}
n_{\rm CR} = 2.46 \times 10^{11} 
\left( \frac{B^2}{100 \, \mathrm{kG}^2} \right)
\left( \frac{1 \, \mathrm{GeV}}{E_{\rm CR}} \right)
\left( \frac{\eta_{\rm eff}}{1} \right).
\end{equation}
If we consider a source with a relatively low, but possible number density of protons, $n_p =10^{7}$ cm$^{-3}$ \cite{Banik:2019twt, Banik:2022nrc, Xue:2022jak}, and fix the cosmic ray energy to $E_{\rm CR} = 10^8$ GeV, the resulting ratio would then be
\begin{equation}
    \frac{n_{\rm CR}}{n_{\rm p}} \sim 2.46\times 10^{-4}\;.
     \label{numberdensityratio}
\end{equation}

Now we want to perform a comparable estimate for the black hole production cross section. For a cosmic ray with energy $E_{\rm CR}=10^{8}$GeV colliding off a thermal proton, $\sqrt{s} \simeq 10^{4}$GeV. Then we get 
\begin{equation}
    \frac{\sigma_{pp \rightarrow \mathrm{BH}+X}(\sqrt{s} = 10^8 \, \mathrm{GeV})}{\sigma_{pp \rightarrow \mathrm{BH}+X}(\sqrt{s} = 10^4\, \mathrm{GeV})} = 10^{\frac{8}{1+n}}\;.
    \label{crosssectionratio}
\end{equation}
For $n = 2$ these two ratios, Eq. \eqref{numberdensityratio} and Eq. \eqref{crosssectionratio} would approximately cancel out, leading to the conclusion that UHECR collisions can actually be relevant for our considerations, provided the source has a strong magnetic field of tens of kG to MG, the acceleration efficiency is of order $\mathcal{O}(0.1-1)$, and the number density of target protons is in the ballpark of $n_p \simeq 10^{7}-10^{9}$ cm$^{-3}$.

An interesting extension of this idea arises in dual and binary supermassive black hole systems. Dual AGNs such as NGC 6240 \cite{Komossa:2002tn} provide environments where two accreting black holes may each sustain distinct acceleration regions. In confirmed SMBH binaries like OJ 287 \cite{Pasumarti:2024igt}, if one black hole powers a jet while the companion drives stochastic acceleration in a turbulent corona, overlapping cosmic-ray populations could naturally enhance cosmic ray–cosmic ray collision rates. Even binaries with two jets, as suggested in sources like PKS 1302–102 \cite{Graham:2015gma} or PG 1302–102 \cite{DOrazio:2015hge}, may facilitate such collisions through misaligned or precessing beams.

In Figure \ref{fig:timescales_UHECR}, we show the cooling timescales induced by cosmic ray-cosmic ray collisions at various values of $M_f$, and their comparison with the timescales induced by cosmic-ray collisions with thermal protons. We assume a source with $n_p=10^{7}$cm$^{-3}$. It can be noticed that each process dominates at different energy ranges, and cosmic-ray-cosmic-ray collisions are therefore not negligible. We also note that there is no hope to probe fundamental scales $M_f \gtrsim 2$ PeV with cosmic ray collisions, since the production timescales are longer than the age of the Universe across all energies at those scales.\\

\emph{\textbf{High-energy neutrino flux from black hole evaporation}}

Until this point in the manuscript, we based our discussion on comparing the different timescales of processes that can occur in the corona of an AGN. This is based on the simple logic that one should expect deviations from standard astrophysical expectations in case such BSM processes occur faster than SM processes. 

Now we go a step further and predict a specific signal of low-scale gravity theories from multi-messenger AGN. In case microscopic black holes are produced at a sizable rate through CR-p and CR-CR collisions, they will populate the corona of the AGN. Because these microscopic black holes have a short lifetime and radiate via Hawking radiation with a concrete temperature $T_H$, one could search for them in the high-energy neutrino and/or gamma-ray spectra.

Let us first calculate the production rate of BH's per unit time and per unit volume for different CR energies. This is given by 
\begin{equation}
    \frac{dN}{dM_{\rm BH}dt dV} = \int_{M_f^2/2m_p}^{\infty} 4\pi \frac{d\phi}{d E_{\rm CR}}n_p \frac{d\sigma}{dM_{\rm BH}} dE_{\rm CR} \;,
    \label{complicatedeventrateformula}
\end{equation}
where $n_p$ is the density of target protons in the corona, and $\phi(E_{\rm CR})$ is the cosmic ray differential spectrum that is determined by the acceleration mechanism of cosmic rays in the AGN. 

In order to estimate how many BH's will be present in the corona and radiate simultaneously, it is crucial how long lived these BH's are. For now, we focus our calculations to the extra-dimensional case and comment on the many species case later. For extra-dimensional theories the lifetime of BH's is 
\begin{equation}
    \tau_{\textrm{extra dim.}}\Big|_{\rm SC}
        = 
            \frac{ C_{\rm ADD} }
            { M_{\textrm{f}} }\mspace{-2mu}
            \left(
                \frac{ M }{ M_{\textrm{f}} }
            \right)^{\!(n+3)/(n+1)}
            ,
            \label{eq:tau-Extra-Dimensions}
\end{equation}
where $C_{\rm ADD}$ is
\begin{equation}
    C_{\textrm{ADD}}
        =
            \frac{ 16\mspace{1mu}\pi^{2}a_{n}^{2} }
            { \alpha\mspace{1.5mu}(n+1)(n+3) }
            \, .
            \label{eq:C-ADD}
\end{equation}

Also necessary for describing the process is the radiation rate which is 
\begin{equation}
    \frac{ d M }{ d t }\bigg|_{\textrm{extra dim.}}
        =
            -\alpha\big( n, T_{\textrm{H}}^{} \big)\,
            T_{\textrm{H}}^{2}
            \; .
            \label{eq:dMdt-Extra-Dimensions}            
\end{equation}
Here $\alpha$ is a numerical coefficient of order one that describes the radiation behaviour of a BH into the thermally accessible species. When $T_H$ is much larger than the mass of the radiated species one can approximate $\alpha$ as a constant value.

\begin{figure}[tb]
\centering
\includegraphics[width=0.5\textwidth]{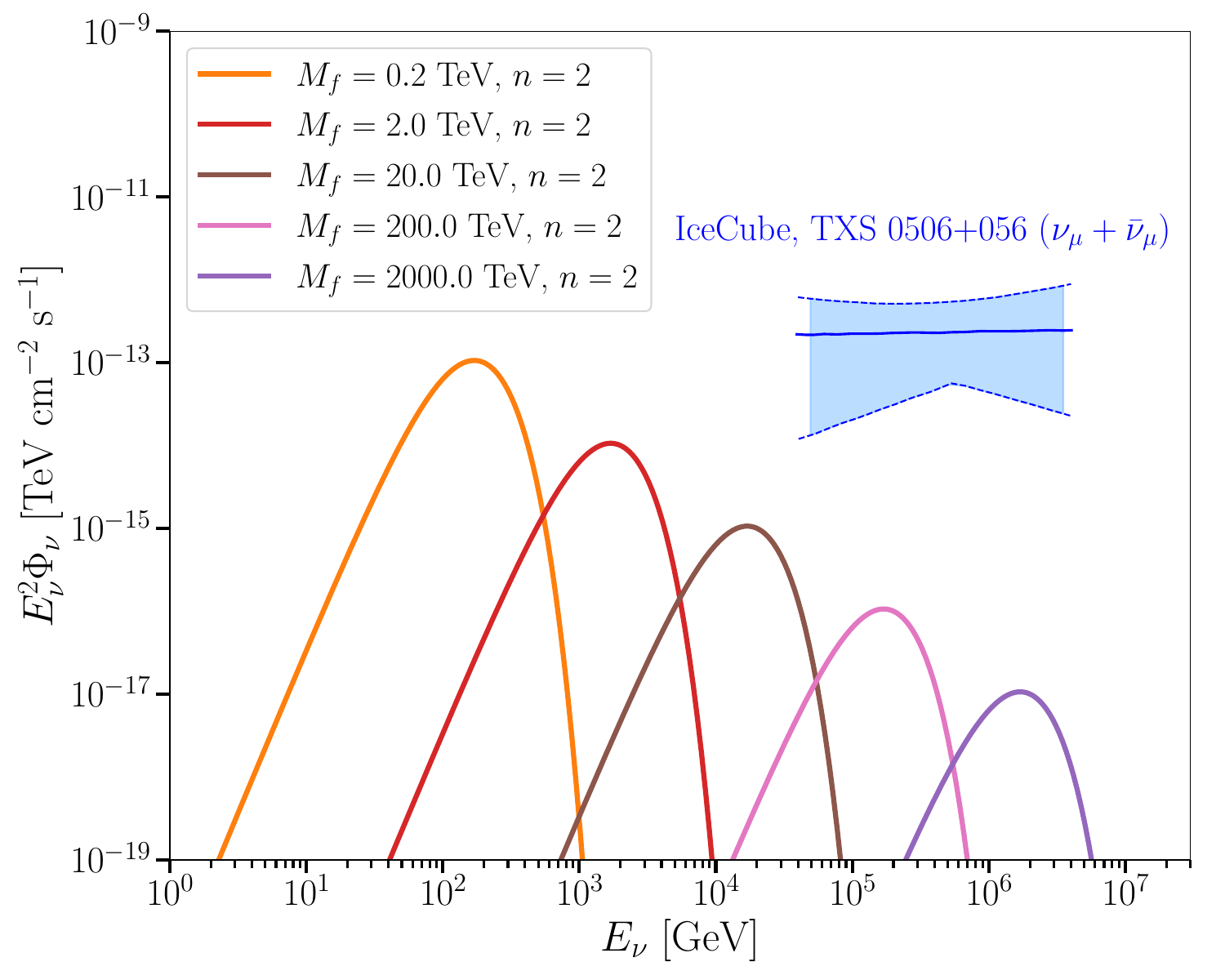}
\caption{High-energy neutrino flux originated from the evaporation of black holes produced in $pp$ collisions at TXS 0506+056, for different values of the fundamental scale of gravity $M_f$, and fixed number of extra-dimensions $n=2$. For comparison, we show as a blue band the TXS 0506+056 high-energy neutrino measurement from the IceCube collaboration \cite{IceCube:2018cha}.}
\label{fig:hawking}
\end{figure}

For a given value of $E_{\rm CR}$, BH's will be dominantly produced of $M_{\rm BH} = \sqrt{s} \simeq \sqrt{2m_p E_{\rm CR}}$, because the geometrical cross section is strictly rising with $E_{\rm CR}$, see \eqref{productioncrosssection}. Knowing the predominant mass of the produced BH's also allows us to calculate their lifetime and Hawking temperature according to \eqref{eq:C-ADD} and \eqref{HawkingTempADD}, respectively. 
Combining this with \eqref{eq:dMdt-Extra-Dimensions} we can predict the number of radiation products of the black holes. Now we arrived at a crucial point. Observe that $T_H$ has an overall factor of $M_f$ and scales with $(M_f/M_{BH})^{1/(1+n)}$ which means the actual temperature changes very slowly with the mass of the black hole and is overall determined by $M_f$. From \ref{tableTH} we see that within the range of $E_{\rm CR}$ over three orders of magnitude, the deviation from $T_H$ just varies by a factor of two. This means that the temperature the signature occurs is determined by $M_f$ and not by $E_{\rm CR}$, which is model dependent.

Notice also that this is a third temperature scale the system is showing. We have the temperature of the thermal protons in the corona, the temperature of the CR's and now the temperature of the radiating BH. Even though these three scales are related they show up on totally different energy scales which enables disentangling the signals. Therefore, a smoking gun signature of micro BH production by AGNs is a black body radiation spectrum showing up in the overall spectrum of the AGN at a temperature of
\begin{equation}
    T_H \approx \frac{M_f(n+1)}{4\pi a_n} \;.
\end{equation}

We show in Figure \ref{fig:hawking} the high-energy neutrino flux arising from micro black hole evaporation in TXS 0506+056, for different values of $M_f$ and fixing $n=2$. For comparison, we show the observed high-energy neutrino flux from TXS 0506+056 by the IceCube collaboration \cite{IceCube:2018cha}. It can be noticed that Hawking radiation yields a thermal spectrum which differs from the power-law astrophysical expectations (and best-fit) in IceCube. The lower the value of $M_f$, the more noticeable the shift compared to astrophysical expectations becomes, and the higher the rate is. The position of the Hawking bump scales linearly with the fundamental scale,
since $E_{\rm peak}\sim \mathcal{O}(1)\times T_H$. By contrast, the overall normalization follows the geometric production rate
$\sigma_{0}\propto r_s^2\propto M_f^{-2}$, such that larger $M_f$ shifts the
bump to higher energy while simultaneously suppressing its amplitude.

\emph{\textbf{High-energy particle emission from different branes.}}\label{brane}
Last but not least we want to point out a unique signature that AGNs can provide us in many brane world scenarios, which are of interest in many extra-dimensional theories. It is usually assumed in extra-dimensional theories that SM particles are confined on a 3+1 dimensional hypersurface called "brane" in a higher dimensional space called the "bulk", where just entities can propagate that are not confined to the brane. Such entities therefore should not carry any SM charges and in particle physics two natural candidates are known namely the graviton and sterile neutrinos. 

So, when there must be one brane in the bulk we are living on, it is a natural extension to think about scenarios with multiple branes that are also populated by particles that could be SM like, or perhaps something unknown to our brane. Prominent examples for such theories of phenomenological relevance are the Randall-Sundrum model \cite{Randall:1999ee}, the manyfold universe \cite{Arkani-Hamed:1999rvc} and $N$naturalness \cite{Arkani-Hamed:2016rle} (see \cite{Langlois:2002bb} for a review on brane cosmology). The actual realization is not so important in our case, but what is crucial is the observation made in \cite{Dvali:2008rm} that macroscopic black holes reach out in the bulk as they are gravitational objects and can accrete the branes that are present in the theory. This means the black hole will accrete and evaporate totally democratically within all the branes that intersect the black hole. 

Now, notice that if on one of these branes the condition is satisfied that an AGN is formed the phenomenology we outlined above takes place and particles like gravitons and sterile neutrinos that can escape into the bulk are produced. These messengers of other branes reach our brane and can decay on it, producing SM particles like gamma rays or high-energy neutrinos with an energy that will be of the order of 
\begin{equation}
    E_{\textrm{messenger}} \lesssim M_f \;,
\end{equation}
as their production is most efficient around such energies. 

In other words, even though the black hole does not have an accretion disk in our brane, it could have one on another brane. Then the black hole would produce messengers whose decay products could reach telescopes based on Earth. This is interesting, as if we ever robustly detect a high-energy neutrino and/or gamma-ray point source arising from a void region without accompanying low-energy multi-wavelength electromagnetic radiation, it could tell us that we are receiving messages of another braneworld.

\begin{figure}[tb]
\centering
\includegraphics[width=0.2\textwidth]{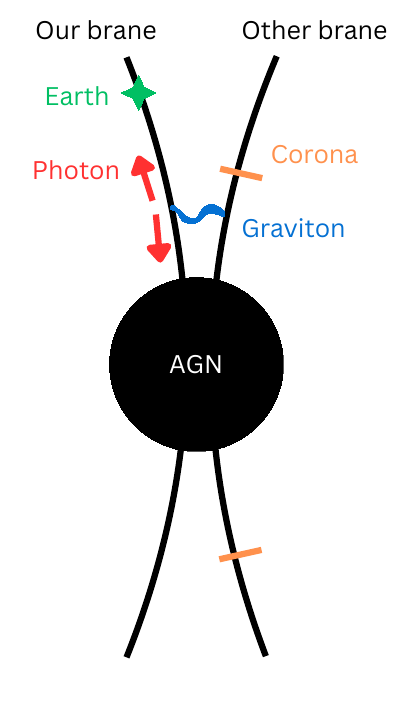}
\caption{Schematic of the possibility that cosmic-ray collisions in an AGN of a different brane produced gravitons that propagate to our brane, potentially yielding high-energy neutrino/photon emission from regions of the sky with no multi-wavelength counterparts.}
\label{fig:brane}
\end{figure}
\emph{\textbf{Conclusions.}}\label{conclusions}
We have investigated what we can learn about the scale of quantum gravity from cosmic ray collisions in multi-messenger Active Galactic Nuclei such as TXS 0506+056 or NGC 1068. We considered both the production of gravitons and black holes via cosmic ray collisions with ambient protons in the AGN, and via cosmic ray collisions with other cosmic rays. We have shown that high-energy neutrino data from TXS 0506+056 allows us to place constraint on the fundamental scale of quantum gravity of $M_f > 0.3$ TeV for $n=2$, with $n$ being the number of extra-dimensions or a model parameter of the many-species theories. We further pointed out that this bound can be pushed to $M_f > 200$ TeV with future high-energy neutrino observations of GZK-neutrinos. We also placed a maximum conceivable sensitivity (upper floor) to the scale of quantum gravity that can be placed with cosmic ray-cosmic ray collisions in AGNs of $M_f \gtrsim 2$ PeV, since the production of black holes for larger scales of quantum gravity would take longer than the Age of the Universe. We showed that cosmic ray-cosmic ray collisions can be relevant in AGN where cosmic rays are accelerated stochastically, or in dual AGN or supermassive black hole binaries.

We also estimated the corresponding high-energy neutrino flux arising from the Hawking evaporation of black holes produced in $pp$ coillisions in TXS 0506+056, finding that the spectra can strongly differ from that expected in standard astrophysical processes such as meson decays. The neutrinos follow a thermal distribution with a peak determined by the scale of quantum gravity $M_f$. This consideration may lead to stronger bounds than those obtained from simply cooling arguments, and we leave this task for future investigation.

Finally, we have speculated with the possibility of detecting isolated high-energy neutrino and/or gamma-ray emission from regions of the sky where no multi-wavelength counterparts are observed. This could occur if gravitons (perhaps produced in $pp$ collisions) produced on a different brane propagate and decay in our Universe. The ensuing high-energy particle emissions would follow a thermal distribution peaking at the scale of Quantum Gravity, different from the typical spectra of observable AGN. Parametrizing and quantifying this potential signature of other Universes is left for future investigation.

Determining the scale of quantum gravity is a longstanding goal of Physics. Naturally, going higher in the maximum testable energy scale is the path to follow, in order to eventually approach the Planck scale. In this work we have discussed a set of novel possibilities involving cosmic ray collisions, and highlighted that current data from TXS 0506+056 already allow us to place constraints at the TeV-scale. Future high-energy neutrino data could open a window to the PeV-scale, and we look forward to that.

\bigskip
{\bf Acknowledgements.} 
We are grateful to the CERN Neutrino Platform for their successful workshop \cite{Abada:2025jpk}, where this projected was initiated. The work of GH is supported by the Neutrino Theory Network Fellowship with contract number 726844, and by the U.S. Department of Energy under award number DE-SC0020262. The work of M.E.~was supported by ANR grant ANR-23-CE31-0024 EUHiggs. This manuscript has been authored by FermiForward Discovery Group, LLC under Contract No.
89243024CSC000002 with the U.S. Department of Energy, Office of Science, Office of High Energy Physics.
 
\bibliography{ref}
\bibliographystyle{utphys} 

\end{document}